\documentclass[a4paper,11pt]{article}

\usepackage{amsmath,amsthm,amsthm,amsfonts}
\usepackage{booktabs,bm,enumitem}
\usepackage[labelsep=period]{caption}
\usepackage[a4paper]{geometry}
\usepackage{setspace}
\usepackage[colorlinks=false, pdfborder={0 0 0}]{hyperref}
\usepackage{rotating}
\usepackage{cases}
\usepackage{authblk}
\usepackage{graphicx}
\allowdisplaybreaks

\usepackage[T1]{fontenc}

\theoremstyle{plain}

\theoremstyle{definition}

\renewcommand{\hat}{\widehat}

\title{On the conditional distribution of the mean of the two closest among a set of three observations}
\author[1]{I.J.H. Visagie}
\author[2]{F. Lombard}
\affil[1]{Department of Statistics, University of Pretoria, South Africa, visagiejaco3@gmail.com}
\affil[2]{Department of Statistics, University of Johannesburg, South Africa}
\date{}

\begin{document}
	\maketitle
	
	\begin{abstract}
	Chemical analyses of raw materials are often repeated in duplicate or
triplicate. The assay values obtained are then combined using a
predetermined formula to obtain an estimate of the true value of the
material of interest. When duplicate observations are obtained, their
average typically serves as an estimate of the true value. On the other
hand, the \textquotedblleft best of three\textquotedblright\ method\
involves taking three measurements and using the average of the two closest
ones as estimate of the true value.

In this paper, we consider another method which potentially involves three
measurements. Initially two measurements are obtained and if their
difference is sufficiently small, their average is taken as estimate of the
true value. However, if the difference is too large then a third independent
measurement is obtained. The estimator is then defined as the average
between the third observation and the one among the first two which is
closest to it.

Our focus in the paper is the conditional distribution of the estimate in
cases where the initial difference is too large. We find that the
conditional distributions are markedly different under the assumption of a
normal distribution and a Laplace distribution.

	\end{abstract}
	
\section{Introduction}

Chemical analyses of raw materials are often repeated in duplicate or
triplicate. The assay values obtained are then combined using a
predetermined formula to obtain an estimate of the true value, $\mu $, of
the material of interest. When duplicate observations $X_{1}$ and $X_{2}$
are obtained, their average typically serves as an estimate of the true
value. On the other hand, the \textquotedblleft best of
three\textquotedblright\ method\ involves taking three measurements $X_{1}$, 
$X_{2}$, and $X_{3}$ and using the average of the two closest of these
values as estimate of the true value. The statistical properties of this
estimator were worked out by Seth (1950) and Lieblein (1952).

In this paper, we consider another method which potentially involves three
measurements. Initially two measurements, $X_{1}$ and $X_{2}$, are obtained.
If the difference between $X_{1}$ and $X_{2}$ is sufficiently small, their
average is taken as the estimate. If the difference is too large, then a
third independent measurement, $X_{3}$, is obtained. Then the estimator,
henceforth denoted by $\hat{\mu}$, is the average between $X_{3}$ and the
one among $X_{1}$ and $X_{2}$ which is closest to $X_{3}$. The rationale
underlying the method is that whichever one of $X_{1}$ and $X_{2}$ is
closest to $X_{3}$ is the least likely to contain a large measurement error.

The usual assumption made in standards documents is that the measurement
error is normally distributed. However, Wilson (1923) draws attention to the
fact that in some instances there are strong grounds for assuming that the
errors follow a Laplace distribution. In the context of a series of
observations that estimate the true value of a given parameter, Keynes
(1911) asks the following question: \textquotedblleft If the most probable
value (maximum likelihood estimate in modern terminology) of the quantity is
equal to the arithmetic mean of the measurements, what law of error does
this imply?\textquotedblright\ Under the additional assumption that the
resulting law of error is symmetric, Keynes shows that it is necessarily
normal. Interestingly, he also shows that when the question is restated to
enquire about the median instead of the mean, then the resulting law of
error is the Laplace distribution which, in standardised form, has density
function%
\begin{equation*}
f\left( x\right) =\frac{1}{\sqrt{2}}\exp \left( -\sqrt{2}\left\vert x-\mu
\right\vert \right) .
\end{equation*}%
These facts provide motivation for studying the behaviour of the estimator, $%
\hat{\mu}$, under both the normal and Laplace distribution assumptions.

Even if both $X_{1}$ and $X_{2}$ are unbiased estimators of $\mu $, the
measurement errors attached to each will result in a fixed proportion $%
\alpha \in \left( 0,1\right) $ of unacceptably large differences. In other
words, a type I error will be made with probability $\alpha $. In this
paper, we investigate the conditional distribution of $\hat{\mu}$ given that
a type I error has occurred. On a purely intuitive level, one would expect
this conditional distribution to be symmetric around $\mu $. This is indeed
the case. However, the form of the symmetry is quite surprising. For
realistic values of $\alpha $ we have the following. It turns out that for
the normal distribution $\hat{\mu}$ has a bimodal conditional distribution
with modes to the left and the right of $\mu $. For the Laplace distribution
the surprise is that $\hat{\mu}$ has a unimodal distribution with mode $\mu $%
.

The remainder of the paper is structured as follows. In Section 2, we define
the estimator and derive its conditional density function in the general
case where $X_{1}$, $X_{2}$ and $X_{3}$ are independent and identically
distribution (i.i.d.) observations from a symmetric distribution. The
conditional density function of the estimator is then computed specifically
in the normal and Laplace cases and the surprising difference between the
two is illustrated and its possible consequences discussed. In Section 3, we
consider a dataset and demonstrate that the Laplace rather than the normal
distribution provides an acceptable fit to the observed data.

\section{Conditional distribution of the estimator}

In the application sketched in the Introduction, the difference between $%
X_{1}$ and $X_{2}$ is regarded as unacceptably large if%
\begin{equation}
\left\vert X_{1}-X_{2}\right\vert >r(\alpha ),  \label{Abs diff criterion}
\end{equation}%
where $r(\alpha )$ satisfies 
\begin{equation}
P\left[ \left\vert X_{1}-X_{2}\right\vert >r(\alpha )\right] =\alpha
\label{definition of r}
\end{equation}%
for an a priori given small positive $\alpha $. In the following, the
argument $\alpha $ in $r\left( \alpha \right) $ is suppressed in cases where
this is unlikely to lead to confusion. Thus, in the absence of any change in
the population mean or standard deviation, the type I error rate will be $%
\alpha $. There are two possibilities, namely

\begin{description}
\item[(i)] $\left\vert X_{1}-X_{2}\right\vert \leq r$, in which case the
estimate $\hat{\mu}=(X_{1}+X_{2})/2$;

\item[(ii)] $\left\vert X_{1}-X_{2}\right\vert >r$, in which case a third
observation $X_{3}$ is obtained and%
\begin{equation}
\hat{\mu}=\left\{ 
\begin{array}{c}
\dfrac{X_{1}+X_{3}}{2}\ if\ |X_{1}-X_{3}|<|X_{2}-X_{3}| \\ 
\\ 
\dfrac{X_{2}+X_{3}}{2}\ if\ |X_{2}-X_{3}|<|X_{1}-X_{3}|.%
\end{array}%
\right.  \label{Reported value}
\end{equation}
\end{description}

Since $\mu $ and the standard deviation of the error distribution, $\sigma $%
, are assumed to be fixed and known, we may assume without loss of
generality that $\mu =0$ and $\sigma =1$.

Our interest centers on (ii), hence on the conditional distribution of $\hat{%
\mu}$ given that $\left\vert X_{1}-X_{2}\right\vert >r$. Let%
\begin{equation}
G\left( x,\alpha \right) =P\left[ \dfrac{X_{1}+X_{3}}{2}\leq x,\
X_{1}-X_{2}>r,\ X_{3}>\dfrac{X_{1}+X_{2}}{2}\right] ,  \label{G def}
\end{equation}%
and 
\begin{equation*}
g\left( x,\alpha \right) =\frac{d}{dx}G\left( x,\alpha \right) .
\end{equation*}%
We show in Appendix 1 that the conditional density function of $\hat{\mu}$,
given $\left\vert X_{1}-X_{2}\right\vert >r$, is%
\begin{equation}
h\left( x,\alpha \right) =\frac{2}{\alpha }\left[ g\left( x,\alpha \right)
+g\left( -x,\alpha \right) \right] .  \label{Xa density}
\end{equation}%
The density $h$ is symmetric around $x=0$ which is what one would expect a
priori. However, from a practitioner's point of view, it is the shape of
this density that turns out to be the most interesting and important aspect
of the conditional distribution. Given a density function $f$ of the $X_{i}$%
, $g\left( x,\alpha \right) $ is given by the expression%
\begin{equation}
g\left( x,\alpha \right) =\int_{-\infty }^{\infty }\int_{-\infty }^{\infty
}2f(2x-x_{1})\mathbb{J}\left( x,x_{1},x_{2}\right) f\left( x_{1}\right)
f\left( x_{2}\right) dx_{1}dx_{2},  \label{Double Integral}
\end{equation}%
where%
\begin{equation}
\mathbb{J}\left( x,x_{1},x_{2}\right) =\mathbb{I}\left( x>\frac{3x_{1}+x_{2}%
}{4},x_{1}-x_{2}>r\right) ,  \label{Def of J}
\end{equation}%
with $\mathbb{I}\left( \cdot \right) $ the indicator function. Substitution
of the normal or Laplace density functions into (\ref{Double Integral}) does
not lead to any substantial algebraic simplification of the expression for $%
h\left( x\right) $. Therefore, we obtain $g(x,\alpha )$ by numerical
integration over a fine grid of $x$ values using the Matlab function
\textquotedblleft integral2.m\textquotedblright\ - see Appendix 2.
    
    \begin{figure}[h!]
        \centering
        \includegraphics[width=.7\textwidth]{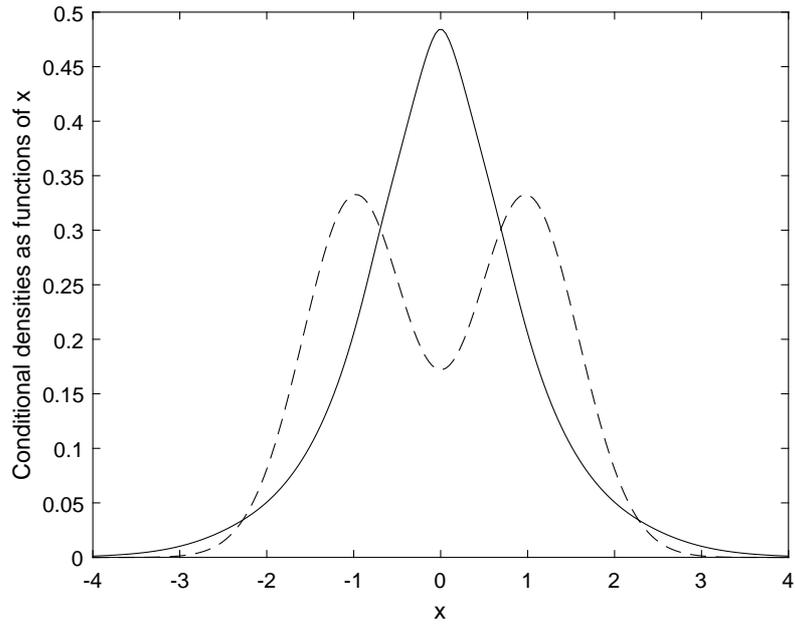}
        \caption{Density of $\hat{%
\protect\mu}$ given that $\left\vert X_{1}-X_{2}\right\vert >r$ under the
normal (dashed line) and Laplace (solid line) assumptions.}
    \end{figure}
    
Figure 1 shows the conditional densities (\ref{Xa density}) of $\hat{\mu}$
in the normal and Laplace distributions. The density in the normal
distribution is bimodal, while in the Laplace distribution it is unimodal.
In both cases, the estimator is centered around the population average.
Nevertheless, a process engineer is bound to be somewhat perplexed upon
seeing the bimodal form in the normal distribution. This phenomenon can, to
some extent, be explained as follows. First, the Laplace distribution
differs from the normal distribution in some important respects. For
instance, the Laplace density has a sharp peak at its point of symmetry,
hence is not differentiable\ there. The tails of the Laplace density are
also substantially thicker than those of the normal density. This is perhaps
not obvious from visual inspection of Figure 2, which shows plots of the
density functions of the two standardised densities.

    \begin{figure}[h!]
        \centering
        \includegraphics[width=.7\textwidth]{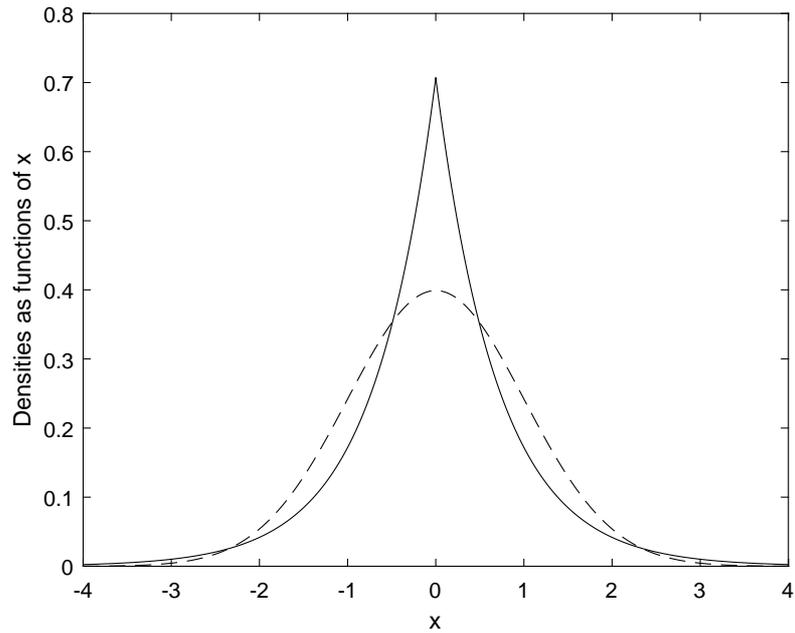}
        \caption{Standardised normal
(dashed line) and Laplace (solid line) density functions.}
    \end{figure}
    
    In order to better appreciate
the differences between the tails of the distributions, consider Table 1,
which shows the numbers $r(\alpha )$\ which make $P\left(
|X_{1}-X_{2}|>r(\alpha )\right) =\alpha $\ for a range of values of $\alpha $%
. The indications are that the Laplace distribution has substantially
heavier tails than the normal distribution. In fact, the kurtosis of the
Laplace distribution is $6$, twice that of the normal distribution.

\begin{center}
\begin{tabular}{|c|c|c|}
\hline
$\alpha $ & $normal$ & $Laplace$ \\ \hline
0.10 & 1.645 & 1.628 \\ \hline
0.05 & 1.960 & 2.118 \\ \hline
0.025 & 2.241 & 2.608 \\ \hline
0.01 & 2.576 & 3.256 \\ \hline
0.005 & 2.807 & 3.746 \\ \hline
\end{tabular}

\begin{tabular}{ll}
Table 1 & Comparison of tail thicknesses of \\ 
& normal and Laplace densities%
\end{tabular}
\end{center}
	
Second, we now argue that, as a consequence of the preceding remark, the
resulting density is bimodal in the case where the separation between $%
g\left( x,\alpha \right) $ and $g\left( -x,\alpha \right) $ per unit
standard deviation is large and unimodal when this separation is small.

Figure 3 shows plots of $g\left( x,\alpha \right) $ and $g\left( -x,\alpha
\right) $ for the normal distribution while Figure 4 shows the corresponding
plots for the Laplace distribution. The figures clearly indicate that the
separation between $g\left( x,\alpha \right) $ and $g\left( -x,\alpha
\right) $ is substantially larger under the normal distribution than under
the Laplace distribution.

    \begin{figure}[h!]
        \centering
        \includegraphics[width=.7\textwidth]{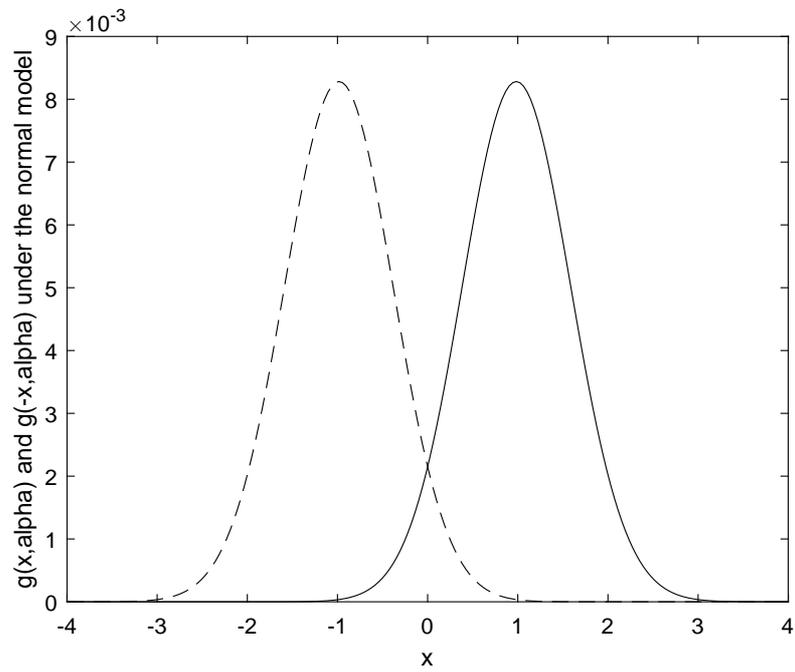}
        \caption{$g\left( x,\protect\alpha \right) $ (solid line) and $g\left( -x,\protect%
\alpha \right) $ (dashed line) under the normal distribution.}
    \end{figure}
    
    \begin{figure}[h!]
        \centering
        \includegraphics[width=.7\textwidth]{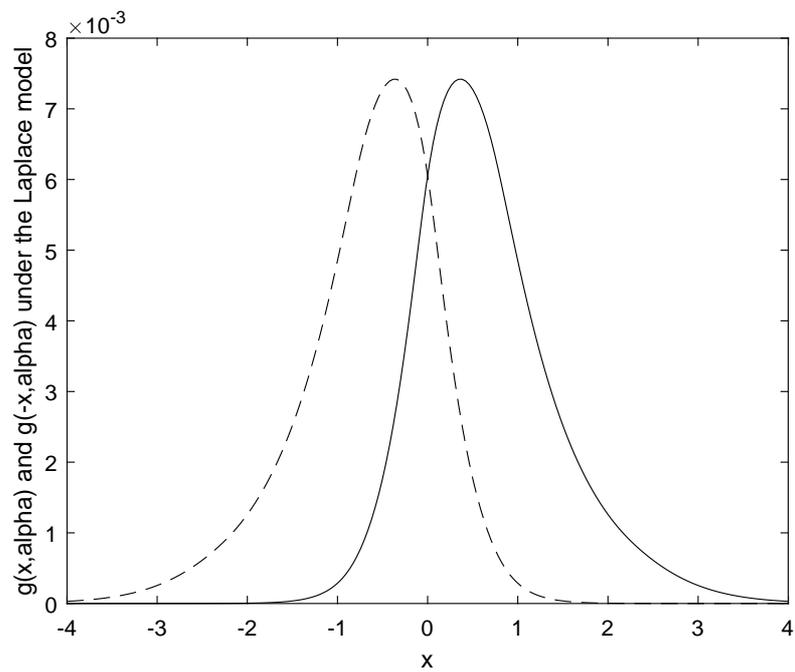}
        \caption{$g\left( x,\protect\alpha \right) $ (solid line) and $g\left( -x,\protect%
\alpha \right) $ (dashed line) under the Laplace distribution.}
    \end{figure}
    
We now discuss some possible consequences of this difference between the two
conditional distributions. The quality of coal is determined, in part, by
its ash content. The lower the ash content, the greater is the release of
energy when the coal is burnt. As a result, the price of coal is often
linked to its ash content. Typically, two determinations, $X_{1}$ and $X_{2}$%
, of the ash content of a batch of coal are made and the estimate, $\hat{\mu}
$, is computed as shown above. As pointed out above, even if both
determinations are unbiased estimators of $\mu $, unacceptably large
deviations would occur in a proportion $\alpha $ of batches. If $\mu $
denotes the contractual ash content, then ash contents in excess of $\mu $
could attract penalties, i.e., a lower price than that originally agreed
upon.

Figure 5 shows conditional exceedance probabilities%
\begin{equation*}
P\left( \hat{\mu}>x\left\vert \left\vert X_{1}-X_{2}\right\vert >r\right.
\right)
\end{equation*}%
over a range of $x$ values for the normal and Laplace distributions.

    \begin{figure}[h!]
        \centering
        \includegraphics[width=.7\textwidth]{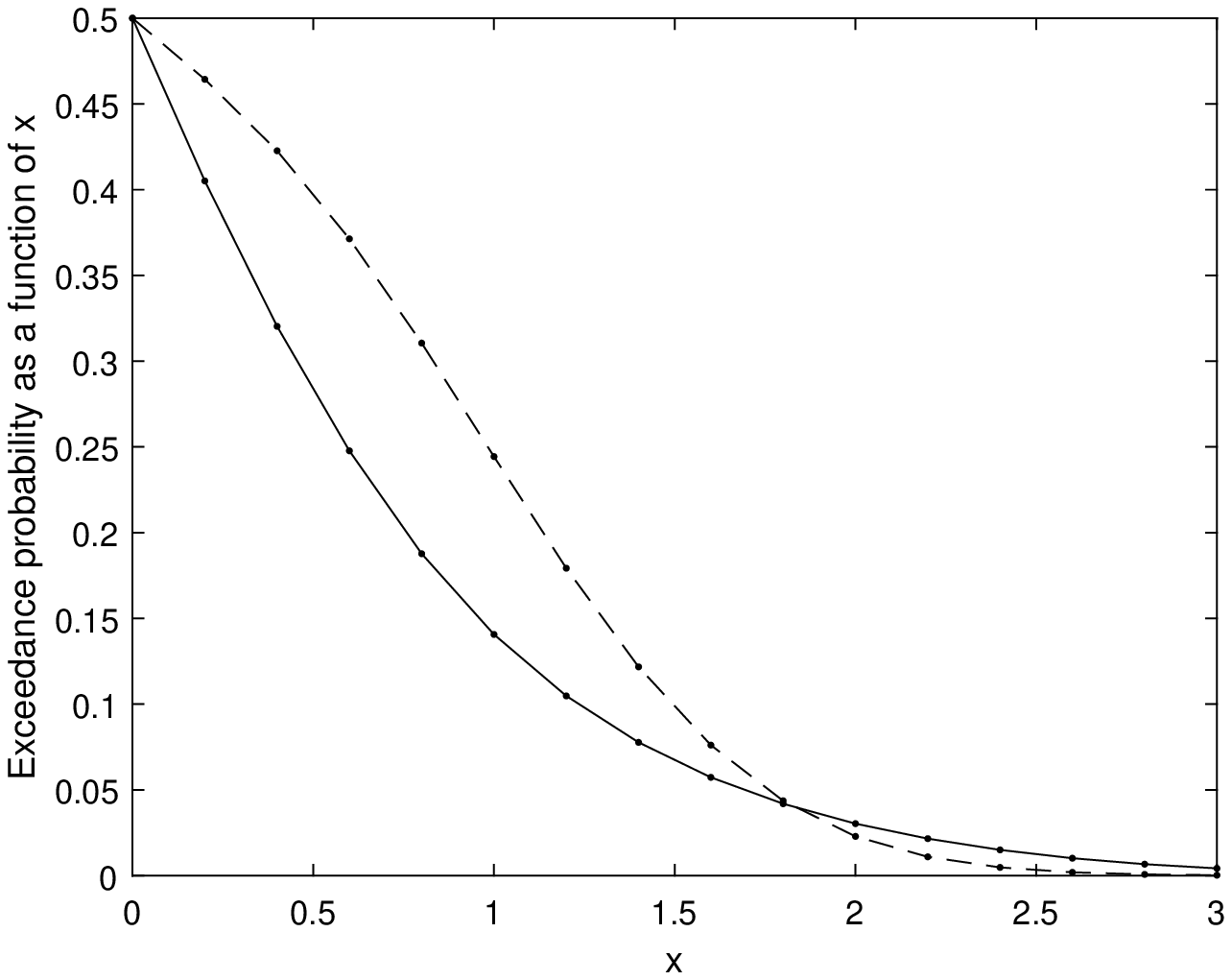}
        \caption{Exceedance probabilities for the normal (dashed line) and Laplace (solid line) distributions.}
    \end{figure}
    
    From the figure it is clear that deviations up to $1.5$ standard deviations
in a normal distribution will tend to attract larger penalties than in a
Laplace distribution. This is also rather clear from Figure 1. The economic
implications of this are greater than would seem to be apparent at first
glance. A batch of coal could consist of several hundreds of tons, which
means that the penalty of, for example, $1\%$ of the contractual price could
involve hundreds of thousands of dollars.

\section{Application to some data}

If an enormous amount of data were available, it would be possible to assess
empirically which of the conditional densities seen in Figure 1 is the valid
one. In the absence of a large amount of data we will have to be satisfied
with something less, namely a test of sorts to decide which of the normal or
Laplace error distributions is applicable. Towards this, Figure 6 shows the
differences $X_{1,j}-X_{2,j}$, $j=1,...,199$, for 199 batches of coal.
Typically, a prescribed value of $\sigma $, the common standard deviation of 
$X_{1}$ and $X_{2}$, is attained by following a standard operating
procedure. In the present instance, the prescribed value was $\sigma =0.4$.
Thus, we standardise the observed differences as follows:%
\begin{equation*}
Z_{j}=\frac{X_{1,j}-X_{2,j}}{0.4}.
\end{equation*}%
The resulting sample mean and standard deviation are $-0.06$ and $1.40\
(\approx \sqrt{2})$ respectively.

    \begin{figure}[h!]
        \centering
        \includegraphics[width=.7\textwidth]{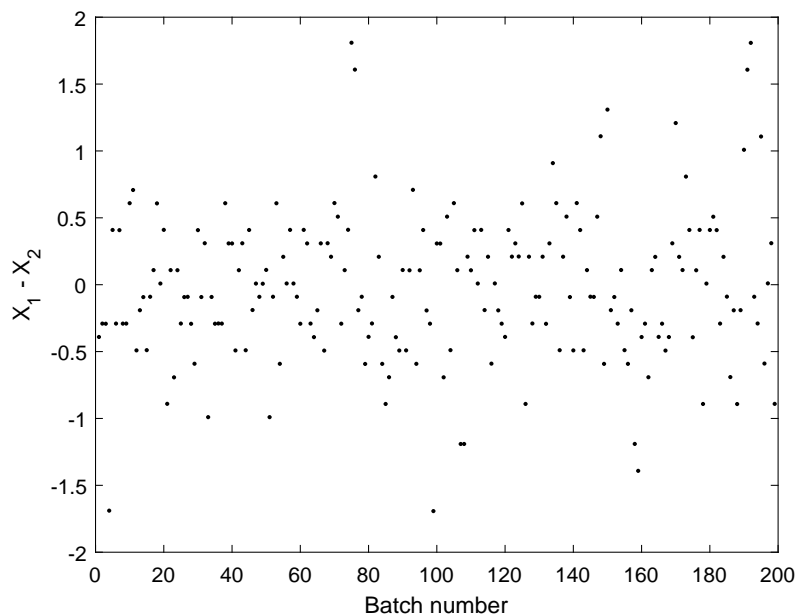}
        \caption{Difference in observed ash contents for 199 batches of coal.}
    \end{figure}
    
    In order to determine which of the two distributions is most appropriate we
use the standardised Kolmogorov-Smirnov statistic:%
\begin{equation}
T_{n}=\max_{1\leq j\leq n}\left\vert \frac{F\left( Z_{j}\right) -F_{n}\left(
Z_{j}\right) }{\sqrt{F\left( Z_{j}\right) \left( 1-F\left( Z_{j}\right)
\right) }}\right\vert _{,}  \label{Tn}
\end{equation}%
where $F$ denotes the cumulative distribution function of $Z$ and $F_{n}$
denotes the usual empirical distribution function%
\begin{equation*}
F_{n}\left( x\right) =\frac{1}{n}\sum_{j=1}^{n}\mathbb{I}\left( Z_{j}\leq
x\right) .
\end{equation*}%
The observed values of $T_{n}$ in the dataset are $T_{n}=0.27$ and $%
T_{n}=0.21$ when $F$ is based on the normal and Laplace error distributions
respectively. The corresponding $p$-values obtained from $100\ 000$ Monte
Carlo simulations are $0.09$ and $0.21$ respectively. These $p$-values
suggest more support for the Laplace assumption than for the normal in this
particular instance.

\section{Appendix 1: Derivation of (\protect\ref{Xa density})}

Let $X_{1}$ and $X_{2}$ denote the first two observations and let $X_{3}$
denote the third sample observation. Given $x$ and a small $\delta >0$, let $%
dx$ denote the interval $(x-\delta ,x+\delta )$. Then%
\begin{eqnarray}
&&P\left[ X\in dx\left\vert |X_{1}-X_{2}|>r\right. \right]  \notag \\
&=&\frac{P\left[ X\in dx,|X_{1}-X_{2}|>r\right] }{P\left[ \left\vert
X_{1}-X_{2}\right\vert >r\right] }  \notag \\
&=&\frac{1}{\alpha }P\left[ X\in dx,|X_{1}-X_{2}|>r\right] .  \label{Eqn 1}
\end{eqnarray}%
Furthermore, since $(X_{1},X_{2},X_{3})$ has the same distribution as $%
(X_{2},X_{1},X_{3})$, 
\begin{equation}
P\left[ X\in dx,|X_{1}-X_{2}|>r\right] =2P\left[ X\in dx,X_{1}-X_{2}>r\right]
.  \label{Eqn 2}
\end{equation}%
Now,%
\begin{eqnarray}
&&P\left[ X\in dx,X_{1}-X_{2}>r\right]  \notag \\
&=&P\left[ \dfrac{X_{1}+X_{3}}{2}\in dx,\ X_{1}-X_{2}>r,\left\vert
X_{1}-X_{3}\right\vert <\left\vert X_{2}-X_{3}\right\vert \right] +  \notag
\\
&&P\left[ \dfrac{X_{2}+X_{3}}{2}\in dx,\ X_{1}-X_{2}>r,\ \left\vert
X_{1}-X_{3}\right\vert >\left\vert X_{2}-X_{3}\right\vert \right]  \notag \\
&=&P\left[ \dfrac{X_{1}+X_{3}}{2}\in dx,\ X_{1}-X_{2}>r,\ X_{3}>\dfrac{%
X_{1}+X_{2}}{2}\right] +  \notag \\
&&P\left[ \dfrac{X_{2}+X_{3}}{2}\in dx,\ X_{1}-X_{2}>r,\ X_{3}<\dfrac{%
X_{1}+X_{2}}{2}\right]  \notag \\
&=&G(x+\delta ,\alpha )-G(x-\delta ,\alpha )+  \label{Eqn 3} \\
&&P\left[ \dfrac{X_{2}+X_{3}}{2}\in dx,\ X_{1}-X_{2}>r,\ X_{3}<\dfrac{%
X_{1}+X_{2}}{2}\right] ,  \notag
\end{eqnarray}%
with the next to last equality following because 
\begin{eqnarray*}
&&X_{1}-X_{2}>r\ and\ \left\vert X_{1}-X_{3}\right\vert <\left\vert
X_{2}-X_{3}\right\vert \\
&\iff &X_{1}>X_{2}+r\ and\ X_{3}\ closer\ to\ X_{1}\mathrm{\ }than\ to\ X_{2}
\\
&\iff &X_{1}>X_{2}+r\ and\ X_{3}>\dfrac{X_{1}+X_{2}}{2}
\end{eqnarray*}%
and%
\begin{eqnarray*}
&&X_{1}-X_{2}>r\ and\ \left\vert X_{1}-X_{3}\right\vert >\left\vert
X_{2}-X_{3}\right\vert \\
&\iff &X_{1}>X_{2}+r\ and\ X_{3}\ closer\ to\ X_{2}\mathrm{\ }than\ to\ X_{1}
\\
&\iff &X_{1}>X_{2}+r\ and\ X_{3}<\dfrac{X_{1}+X_{2}}{2}_{.}
\end{eqnarray*}%
Next, the second term in (\ref{Eqn 3}) is

\begin{eqnarray}
&&P\left[ \dfrac{X_{2}+X_{3}}{2}\in dx,\ X_{1}-X_{2}>r,\ X_{3}<\dfrac{%
X_{1}+X_{2}}{2}\right]  \notag \\
&=&P\left[ \dfrac{-X_{2}-X_{3}}{2}\in d(-x),\ (-X_{2})-(-X_{1})>r,\ -X_{3}>%
\dfrac{-X_{1}-X_{2}}{2}\right]  \notag \\
&=&P\left[ \dfrac{X_{1}+X_{3}}{2}\in d(-x),\ X_{1}-X_{2}>r,\ X_{3}>\dfrac{%
X_{1}+X_{2}}{2}\right]  \notag \\
&=&G(-x+\delta ,\alpha )-G(-x-\delta ,\alpha ),  \label{Eqn 4}
\end{eqnarray}%
with the next to last equality following because $(-X_{2},-X_{1},-X_{3})$
has the same distribution as $(X_{1},X_{2},X_{3})$. Putting (\ref{Eqn 1}), (%
\ref{Eqn 2}), (\ref{Eqn 3}) and (\ref{Eqn 4}) together, we see that%
\begin{eqnarray*}
&&\frac{P\left[ X\in dx\left\vert |X_{1}-X_{2}|>r\right. \right] }{2\delta }
\\
&=&\frac{G(x+\delta ,\alpha )-G(x-\delta ,\alpha )}{2\delta }+\frac{%
G(-x+\delta ,\alpha )-G(-x-\delta ,\alpha )}{2\delta }_{.}
\end{eqnarray*}%
Letting $\delta \downarrow 0$ gets us to (\ref{Xa density}):%
\begin{equation*}
P\left[ X\in dx\left\vert |X_{1}-X_{2}|>r\right. \right] =\frac{2}{\alpha }%
\left( g(x,\alpha )+g(-x,\alpha )\right) .
\end{equation*}

\section{Appendix 2: Derivation of (\protect\ref{Double Integral})}

Let $X_{1}$, $X_{2}$ and $X_{3}$ be independent random variables with common
distribution function $F$ and density function $f$. Then, for fixed $x_{1}$
and $x_{2}$,%
\begin{eqnarray*}
&&P\left( \left. \frac{X_{1}+X_{3}}{2}\leq x,X_{1}-X_{2}>r,X_{3}>\frac{%
X_{1}+X_{2}}{2}\right\vert X_{1}=x_{1},\ X_{2}=x_{2}\right) \\
&=&P\left( \frac{x_{1}+X_{3}}{2}\leq x,X_{3}>\frac{x_{1}+x_{2}}{2}\right) 
\mathbb{I}\left( x_{1}-x_{2}>r\right) \\
&=&P\left( \frac{x_{1}+x_{2}}{2}<X_{3}<2x-x_{1}\right) \mathbb{I}\left(
2x-x_{1}>\frac{x_{1}+x_{2}}{2},\ x_{1}-x_{2}>r\right) \\
&=&\left( F\left( 2x-x_{1}\right) -F\left( \frac{x_{1}+x_{2}}{2}\right)
\right) \mathbb{J(}x,x_{1},x_{2}),
\end{eqnarray*}%
where $\mathbb{J}$ is defined in (\ref{Def of J}). Consequently,%
\begin{eqnarray*}
G\left( x,a\right) &=&E\left[ P\left( \left. \frac{X_{1}+X_{3}}{2}\leq x,\
X_{1}-X_{2}>r,\ X_{3}>\frac{X_{1}+X_{2}}{2}\right\vert X_{1},X_{2}\right) %
\right] \\
&=&E\left[ \left( F\left( 2x-X_{1}\right) -F\left( \frac{X_{1}+X_{2}}{2}%
\right) \right) \mathbb{J(}x,X_{1},X_{2})\right] .
\end{eqnarray*}%
Taking the derivative with respect to $x$, we obtain%
\begin{eqnarray*}
g\left( x,\alpha \right) &=&E\left[ 2f\left( 2x-X_{1}\right) \mathbb{J(}%
x,X_{1},X_{2})\right] \\
&=&\int_{-\infty }^{\infty }\int_{-\infty }^{\infty }2f(2x-x_{1})\mathbb{J}%
\left( x,x_{1},x_{2}\right) f\left( x_{1}\right) f\left( x_{2}\right)
dx_{1}dx_{2}.
\end{eqnarray*}
	

\begin{thebibliography}{9}
\bibitem{} Keynes, J.M. (1911). The principal averages and the laws of error
which lead to them. \textit{Journal of the Royal Statistical Society,} 
\textbf{74, }322-331.

\bibitem{} Lieblein, J. (1952). Properties of certain statistics involving
the closest pair in a sample of three observations\textit{,} \textit{Journal
of Research of the National Bureau of Standards,} \textbf{48} (3)\textbf{, }%
255-268.

\bibitem{} MATLAB Release 2018b, The MathWorks, Inc., Natick, Massachusetts,
United States.

\bibitem{} Seth, G.R. (1950). On the distribution of the two closest among a
set of three observations. \textit{The Annals of Mathematical Statistics,} 
\textbf{21}(2), 298-301.

\bibitem{Wilson} Wilson, E.B. (1923). First and second laws of error. 
\textit{Journal of the American Statistical Association}, \textbf{18},
841-851.
\end{thebibliography}
\end{document}